\documentstyle[twoside,fleqn,espcrc2,psfig,cite]{article}

\def\simgt{\rlap{\lower 3.5 pt \hbox{$\mathchar \sim$}}%
           \raise 1pt \hbox {$>$}}
\def\simlt{\rlap{\lower 3.5 pt \hbox{$\mathchar \sim$}}%
           \raise 1pt \hbox {$<$}}

\newcommand{\AmS}{{\protect\the\textfont2
  A\kern-.1667em\lower.5ex\hbox{M}\kern-.125emS}}

\title{New Developments in Lattice QCD:
Calculation of Flavor Singlet Nucleon Matrix Elements
and Hadron Scattering Lengths}

\author{Masanori Okawa\address{National Laboratory for High Energy
Physics (KEK), Tsukuba, Ibaraki 305, Japan}}

\begin{document}

\begin{abstract}
Recent developments in lattice QCD calculation of flavor singlet nucleon
matrix elements are reviewed.
Substantial sea quark contributions are found in the $\pi$-$N\ \sigma$
term and the quark spin content of the nucleon such that the total
magnitude including valence contributions is in reasonable agreement
with experiments.
Some problems with flavor non-singlet nucleon matrix elements are pointed
out.  Recent work on lattice QCD calculation of hadron scattering length
is also discussed.
\end{abstract}

\maketitle

\section{INTRODUCTION}

While extensive effort has been invested for studies of the flavor
non-singlet sector of strong interactions with the use of lattice QCD,
attempts toward understanding the flavor singlet sector were quite limited
until a few years ago\cite{fku,iiy}.  This stemmed from the severe
technical difficulty that a calculation  of disconnected quark loop
amplitudes, not present in the flavor non-singlet sector, requires quark
propagators for a number of source points equal to the space-time lattice
volume, which is prohibitively computer time consuming if it is to be
carried out with the conventional method of point source.

Several methods have been developed
in the last few years, however, for overcoming this problem.
One of the method is a variant
version of the wall source technique in which gauge configurations are not
fixed to any gauge\cite{pipisct}, as was employed in the early studies of
extended sources\cite{kenway}.  Another is an improvement of the method of
noisy source\cite{nsource} with the use of a random $Z_2$
noise\cite{Z2source}.

The first method has been successfully
applied to estimate the flavor singlet
$\eta'$ meson mass in quenched QCD\cite{refeta}.
The method also allows a calculation of full hadron four-point functions,
for which a similar computational problem is present, and hence an
evaluation of hadron scattering lengths\cite{pipisct,NNsct,hadsct}.
These results were available at the time of the Dallas
Conference\cite{lat93}.

In this article we review new developments in this field since then.
We devote the major part of the review to discuss recent calculations of
flavor singlet nucleon matrix
elements\cite{lat94,ktsigma,ktspin,liulat,liuspin,altsigma,altspin,guptaB,
guptaM}, specifically
the $\pi$-$N$ $\sigma$ term and the axial vector matrix elements (quark
content of proton spin).  We also point out some problems with flavor
non-singlet nucleon matrix elements\cite{lat94}
which were not apparent in the
pioneering studies\cite{romeold,wuppertalold}.
In the rest of the article we turn to some recent work on hadron
scattering lengths\cite{BG,hadsct}.

Most of calculations discussed in this review are carried out in quenched
QCD with the Wilson quark action.  This should be understood unless stated
otherwise.

\section{CALCULATIONAL TECHNIQUES}

\begin{figure}[t]
\vspace*{-0.8cm}
\begin{center}
\leavevmode\psfig{file=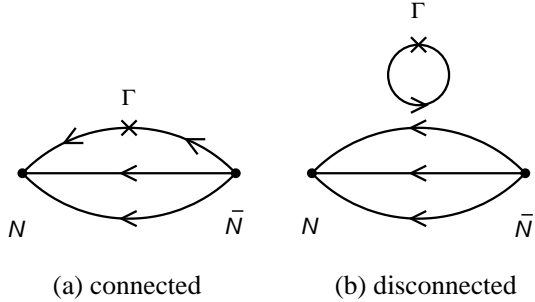,height=4.0cm,width=7.0cm,angle=-90}
\end{center}
\vskip -1.2cm
\caption{(a) Connected and (b) disconnected contribution to the
three-point function of nucleon and flavor singlet operator
$\bar q\Gamma q$.}
\label{fig:threep}
\vspace{-6mm}
\end{figure}

A standard method for extracting the nucleon matrix element of an
operator $O$ is to
employ the formula for
the ratio of three- to two-point function of the nucleon and the
operator given by\cite{romeold}
\begin{eqnarray}
R(t)&=& \frac{\langle N(t) \sum_{n} O(n)
{\bar N}(0)\rangle}
     {\langle N(t){\bar N}(0)\rangle} \nonumber \\
    &\stackrel{t\gg 1}{\longrightarrow}&
     {\rm const}\ +\
     Z_O^{-1}\langle N\vert O \vert N\rangle\ t,
\label{eq:ratio}
\end{eqnarray}
where $N(t)$ is the nucleon operator projected to zero momentum and
$Z_O$ is the lattice renormalization factor for the operator $O$.
For a flavor singlet quark bilinear operator $O=\overline{q}\Gamma q$,
there are two types of diagrams contributing to the three-point
function shown in Fig.~\ref{fig:threep}.
The connected amplitude can be calculated by the
well known source technique\cite{source}.
To handle the disconnected piece,
for which the source point of the disconnected quark loop has to be
summed over
all space-time sites, two methods have been developed.

\subsection{Wall source method without gauge fixing}

In this method one calculates quark propagators with unit source at every
space-time site without gauge fixing\cite{pipisct,lat94}.
The result $\hat G(n)$ is a sum of point-to-point propagators of
form
\begin{equation}
{\hat G}(n)\
=\ \sum_{n'} G(n;n').
\label{eq:ghat}
\end{equation}
The product of the nucleon propagator and
$\sum_n{\rm Tr}[{\hat G}(n)\Gamma]$
equals the disconnected amplitude up to gauge-variant nonlocal terms,
which, however, cancel in the average over gauge configurations.
Generalizing the well-known argument\cite{error}, it is possible
to show that the magnitude of gauge-variant noise terms is smaller than
the signal by a factor
$1/\sqrt{VN}$ for a  sufficiently large volume $V$ and a number of
configurations $N$.
In practice $O(100-200)$ configurations are employed to obtain
reliable signals on a lattice such as $16^3\times 20$ at
$\beta=5.7$\cite{ktsigma,ktspin}.

This method requires only two quark
matrix inversions for each gauge configuration, one for calculating the
nucleon propagator and the other for the disconnected quark loop.
Repeating the latter $n$ times, applying
random gauge transformations to a given gauge field, does not yield much
gain since errors are reduced only by a factor $1/\sqrt{n}$
while the computer time grows proportional to $n$.

\begin{table*}[t]
\vspace*{-0.5cm}
\setlength{\tabcolsep}{1.5pc}
\newlength{\digitwidth} \settowidth{\digitwidth}{\rm 0}
\catcode`?=\active \def?{\kern\digitwidth}
\caption{Recent runs for nucleon matrix elements.
W and KS refer to Wilson and Kogut-Susskind quarks.}
\label{tab:effluents}
\begin{tabular*}{\textwidth}{@{}l@{\extracolsep{\fill}}llllll}
\hline
  & $\beta$ & $m_\pi/m_\rho$ & size & $\#$ conf. \\
\hline
singlet matrix elements\\[1mm]
Kyoto-Tsukuba\cite{lat94,ktsigma,ktspin}$\ \ \ \ \ $
& 5.7 (quenched W) &0.6 - 0.86 & $16^3\times20$ & 260 \\
Kentucky\cite{liulat,liuspin}
& 6.0 (quenched W) & 0.8 - 0.95 & $16^3\times24 $ & 50 \\
Altmeyer et al.\cite{altspin}
& 5.35 ($N_f=4$ KS) & 0.46 & $16^3\times24 $ & 85 \\
Gupta et al. full\cite{guptaB}
&5.4-5.6 ($N_f=2$ W)&0.7 - 0.85
&$16^3\times(16 \times 2)$&15 - 45\\[2mm]
non-singlet matrix elements\\
Gupta et al.\cite{guptaB}
& 6.0 (quenched W) & 0.72 - 0.79 & $16^3\times40 $ & 35 \\
Liu et al.\cite{liuaxial}
& 6.0 (quenched W) & 0.8 - 0.95 & $16^3\times24 $ & 24 \\
G${\rm \ddot o}$ckeler et al.\cite{desy}
& 6.0 (quenched W) & 0.7 - 0.89 & $16^3\times32 $ & 400-1000 \\
\hline
\end{tabular*}
\label{table:tabone}
\end{table*}

\subsection{$Z_2$ source method}

In this method\cite{Z2source} $L$ quark propagators $G^{\ell}_n$ are
calculated for an $L$
set of random $Z_2$ source $\eta^{\ell}_n$ placed at every space-time
site for each gauge
configuration, {\it i.e.,}
\begin{equation}
G^{\ell}_{n}\
=\ \sum_{n'} G(n;n')\eta^{\ell}_{n'},\ \ 1 \le \ell \le L.
\label{eq:gell}
\end{equation}
The product of the nucleon propagator and
${1 \over L}\sum_{\ell\, n}{\rm Tr}[G_{n}^{\ell}\eta_{n}^{\ell}\Gamma]$
equals the disconnected amplitude since
$\lim_{L\to\infty}{1 \over L}\sum_{\ell=1}^L \eta_n^{\ell}\eta_{n'}^{\ell}
= \delta_{nn'}$.

Employing a random source for evaluating disconnected amplitudes
was previously tried for the $\pi$-$N$ $\sigma$ term for the
Kogut-Susskind quark action without much success\cite{altsigma}.
The improvement with the $Z_2$ noise is that
the variance of the result for a finite $L$ is
smallest compared to other types of noise
including the Gaussian noise.

The $Z_2$ source method allows a calculation of disconnected amplitudes
for each configuration.  However,
concrete applications of the method\cite{liulat,liuspin}
show that the number of quark propagators $L$ has to be
large, generally exceeding 100.  Thus the method becomes harder to use
for lighter quarks for which a substantial computer time is
necessary to obtain even a single quark propagator.

\subsection{Other methods}

Some flavor singlet nucleon matrix elements can be estimated indirectly
employing their relation with other quantities: (i)
the scalar density matrix element can be related to the nucleon mass
via Hellmann-Feynman theorem,
\begin{equation}
\langle N \vert {\bar q} q \vert N \rangle\ =\ \frac{dm_N}{dm_q},
\label{eq:hellmann}
\end{equation}
(ii) for the axial vector matrix element the $U(1)$ anomaly
equation yields
\begin{eqnarray}
&&\langle {\vec p},s \vert {\bar q}\gamma_\mu \gamma_5 q
\vert {\vec p},s \rangle s_\mu \nonumber \\
&\propto&\lim_{{\vec q} \to 0}
{1 \over {\vec q}\cdot {\vec s}}
\langle {\vec p}+{\vec q},s \vert {\rm Tr} F_{\mu\nu}{\tilde F}_{\mu\nu}
\vert {\vec p},s \rangle.
\label{eq:anomaly}
\end{eqnarray}
Application of these relations require full QCD simulations if
disconnected contributions
are to be included.  In particular the anomaly relation
(\ref{eq:anomaly}) fails in quenched QCD due to the double pole
in the $\eta'$ meson propagator connecting the disconnected quark loop
and the nucleon\cite{guptaM}.
For these reasons, only a few attempts have been
made with these methods in the past\cite{guptaB,altspin}.

\section{NUCLEON SCALAR MATRIX ELEMENTS}

\subsection{$\pi$-$N$ $\sigma$ term -- phenomenology}

The $\pi$-$N\ \sigma$ term is defined as the product of
the nucleon matrix element of the scalar density and the average
quark mass $\hat m=(m_u+m_d)/2$
\begin{equation}
\sigma(t)\ =\ \hat m
\langle N(p') \vert {\bar u} u + {\bar d} d \vert N(p) \rangle
\label{eq:sigma}
\end{equation}
evaluated at $t$ = $(p'-p)^2$ = 0, {\it i.e.,} $\sigma\equiv\sigma(0)$.
Current algebra and PCAC relate $\sigma(t)$ to the crossing even
$\pi$-$N$ scattering amplitude at the unphysical Cheng-Dashen point
$t=2m_\pi^2$ and a dispersion analysis of $\pi$-$N$ scattering leads to
the value $\sigma(2m_\pi^2)=64(8)$MeV\cite{koch,update}.

On the other hand one can write
\begin{equation}
\sigma(0)\ =\ {\sigma_0 \over 1-y},\quad
\sigma_0=\hat m\langle N\vert\overline{u}u+\overline{d}d-2\overline{s}s
\vert N\rangle
\label{eq:sigma0}
\end{equation}
with
\begin{equation}
y=\frac{2\langle N\vert {\bar s} s \vert N \rangle}
{\langle N\vert {\bar u} u + {\bar d} d \vert N \rangle}.
\end{equation}
Treating flavor $SU(3)$ breaking to first order, one can
estimate\cite{langacker}
\begin{equation}
\sigma_0\approx\frac{\hat m}{m_s-\hat m}\left(M_\Xi+M_\Sigma-2M_N\right)
\approx 25\mbox{MeV},
\label{eq:firstorder}
\end{equation}
where $m_s$ is the strange quark mass.
If one assumes that the variation
$\Delta_\sigma = \sigma(2m_\pi^2) - \sigma(0)$
is small,
the above two estimations lead to a large strangeness content for the
nucleon $y\approx 0.6$.

It has been argued, however, that the variation is substantial,
$\Delta_\sigma \approx$ 15 MeV\cite{update},
in which case the value of the $\sigma$ term
is reduced to $\sigma\approx$ 45 MeV.
Combined with the suggestion\cite{gasser} that one-loop chiral perturbative
corrections raise the value of $\sigma_0$ to $\sigma_0\approx 35$MeV,
this implies a more reasonable value $y\approx 0.2$ for the nucleon
strangeness content.

Let us comment that the magnitude of variation $\Delta_\sigma$ may be
examined within lattice QCD by extrapolating $\sigma(t)$ from the physical
region $t < 0$.  A first attempt has been reported at this
conference\cite{deltasigma}.

\subsection{$\pi$-$N$ $\sigma$ term -- recent lattice results}

In Table~\ref{table:tabone} we list recent studies of nucleon matrix
elements together with parameters of runs.

\begin{figure}[t]
\vspace*{-0.8cm}
\begin{center}
\leavevmode\psfig{file=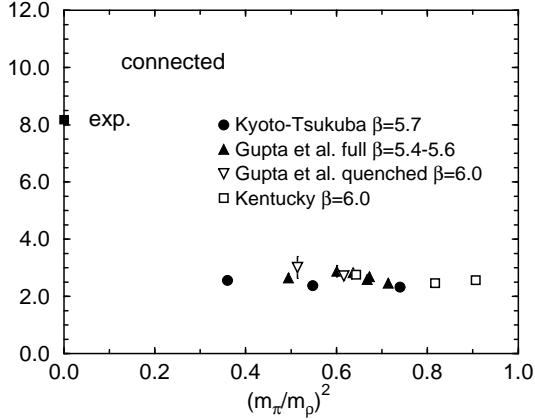,height=5.6cm,width=7.0cm,angle=-90}
\end{center}
\vskip -1.2cm
\caption{Connected contribution
$\langle N\vert {\bar u} u + {\bar d} d \vert N \rangle_{\rm conn.}$
for the scalar matrix element as a function of $(m_\pi/m_\rho)^2$.
For reference see Table 1.}
\label{fig:sigma_conn}
\vspace{-7mm}
\end{figure}

Results obtained for the connected contribution in the scalar matrix
element $\langle N\vert {\bar u} u + {\bar d} d \vert N
\rangle_{\rm conn.}$ are
plotted in Fig.~\ref{fig:sigma_conn}, where we used the
tadpole-improved renormalization factor in the ${\overline {\rm MS}}$
scheme at the scale $\mu=1/a$ given by\cite{tadpole}
\begin{equation}
Z_S\ =\ \left(1-{3K \over 4K_c}\right)
\biggl[1-0.0098\alpha_{\overline {\rm MS}}({1 \over a})\biggr].
\label{eq:Zs}
\end{equation}
The \lq\lq experimental'' value at $(m_\pi/m_\rho)^2=0$ equals
$\langle N\vert {\bar u} u + {\bar d} d \vert N \rangle =\sigma/{\hat m}$
calculated with $\sigma=45$ MeV and ${\hat m}=5.5$ MeV.
Clearly the connected contribution
is too small to account for the experimental value of the
$\pi$-$N\ \sigma$ term.

Previous results for the connected
contribution\cite{romeold,wuppertalold}
are consistent with those in
Fig.~\ref{fig:sigma_conn}.
Estimates from the nucleon mass in quenched QCD  using
the relation (4) also give consistent values.

\begin{figure}
\vspace*{-1.5cm}
\begin{center}
\leavevmode\psfig{file=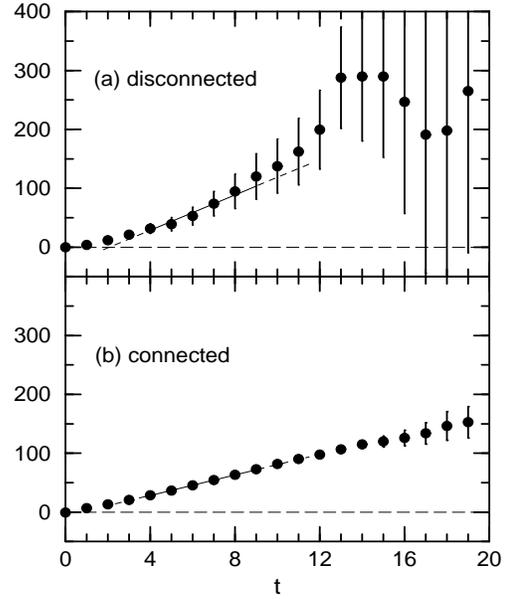,height=8.0cm,width=6.5cm}
\end{center}
\vskip -1.2cm
\caption{Ratio $R(t)$ for the (a) disconnected and (b) connected
amplitudes of the scalar matrix element
$\langle N\vert {\bar u} u + {\bar d} d \vert N \rangle$
evaluated by the wall source method without gauge
fixing\cite{lat94,ktsigma}
for $K$=0.164 at $\beta$=5.7
on a $16^3\times 20$ lattice.
Solid lines are linear fits over $4 \le t \le 9$.}
\label{fig:rskt}
\vspace{-6mm}
\end{figure}

\begin{figure}
\vspace*{-0.8cm}
\begin{center}
\leavevmode\psfig{file=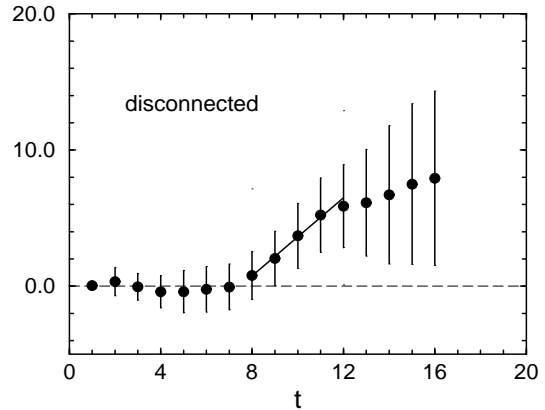,height=5.5cm,width=7.0cm,angle=-90}
\end{center}
\vskip -1.2cm
\caption{Ratio $R(t)$ for the disconnected amplitude
of the scalar matrix element
$\langle N\vert {\bar u} u  \vert N \rangle$
evaluated by the $Z_2$ source method\cite{liulat,liuspin}
for $K$=0.154 at $\beta$=6.0
on a $16^3\times 24$ lattice.  Solid line is a linear fit over
$8 \le t \le 12$.}
\label{fig:rsliu}
\vspace{-6mm}
\end{figure}

A direct calculation of the disconnected amplitude
(Fig.~\ref{fig:threep}(b)) has been made in quenched QCD
using both the wall source method without gauge fixing\cite{lat94,ktsigma}
and the $Z_2$ source method\cite{liulat,liuspin}.
In Fig.~\ref{fig:rskt} results for
the $R$ ratio (\ref{eq:ratio}) obtained with the wall
method\cite{lat94,ktsigma}
at $\beta=5.7$ and $K=0.164$ $(m_\pi/m_\rho=0.74)$ are
plotted.
We observe reasonable signals with a
linear increase in $t$ for the disconnected
amplitude(Fig.~\ref{fig:rskt}(a)), albeit errors are
significantly larger compared with those for the connected
amplitude(Fig.~\ref{fig:rskt}(b)).

The disconnected amplitude obtained with the $Z_2$ source
\cite{liulat,liuspin}
at $\beta=6.0$ and
$K=0.154$ $(m_\pi/m_\rho=0.80)$ is shown in
Fig.~\ref{fig:rsliu}.
A reasonable linear increase is also seen for this case.

\begin{figure}[t]
\vspace*{-0.8cm}
\begin{center}
\leavevmode\psfig{file=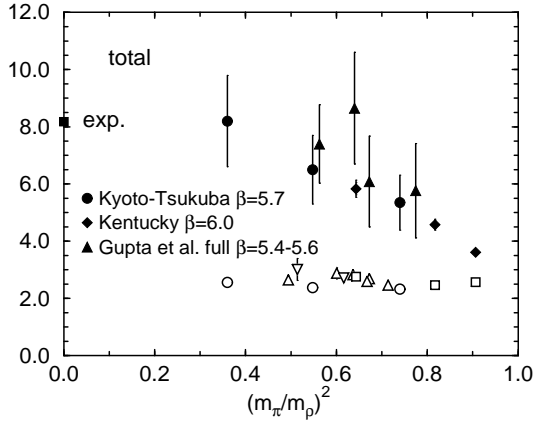,height=5.6cm,width=7.0cm,angle=-90}
\end{center}
\vskip -1.2cm
\caption{Total amplitude of the scalar matrix element
$\langle N\vert {\bar u} u + {\bar d} d \vert N \rangle$
as a function of $(m_\pi/m_\rho)^2$.  For reference see Table 1.
Triangles are full QCD results obtained with the relation (4).}
\label{fig:sigma_total}
\vspace{-6mm}
\end{figure}

An important outcome of these calculations is that the disconnected
contribution to the scalar matrix element is large,
as is clear from a similar magnitude of
slope in Fig.~\ref{fig:rskt} (a) and (b).
Furthermore the total magnitude of the matrix element obtained by adding
the disconnected contribution to the connected one is comparable to the
phenomenological estimate.  This is shown in Fig.~\ref{fig:sigma_total}
where solid symbols represent total values and open ones the connected
contribution
duplicated from Fig.~\ref{fig:sigma_conn}.
A linear extrapolation of quenched results to the chiral
limit yields values of the matrix element tabulated in Table~2 (a).

The filled triangles in Fig.~\ref{fig:sigma_total} represent data obtained
in two-flavor full QCD using the nucleon mass derivative
$dm_N/dm_q$\cite{guptaB}, showing an agreement with quenched estimates
(filled circles and diamonds) in the region of heavy quark with
$m_\pi/m_\rho >0.7-0.8$.

We find these results to be quite encouraging.  On a closer inspection,
however, we notice several problems, to which we now turn.

\subsection{Problems with scalar matrix elements}

\subsubsection{physical value of the $\pi$-$N$ $\sigma$ term and quark
mass}

To convert results for the scalar matrix elements into those
for the $\sigma$ term, we need the value of the
quark mass ${\hat m}$ in physical units\cite{qmassR}.
In quenched QCD current lattice
estimates are $\hat m=5-6$MeV for the Wilson quark action,
in an agreement with the
phenomenological estimation of
${\hat m}=$5.5MeV.  Combined with the total values of the
scalar matrix element in Table~2 (a),
quenched lattice results for the $\sigma$ term
are consistent with the phenomenological value  $\sigma \approx 45$ MeV.

On the other hand, full QCD simulations with the Wilson quark
action yield
${\hat m}\approx 2-3$MeV (see {\it e.g.,} Ref.~\cite{guptaB}),
which are a factor $2-3$ smaller than in the
quenched case.  Since values of matrix elements are similar between
the two cases, this means that the $\sigma$ term is also
small $\sigma\approx 20$MeV in full QCD.

Let us add that the $\sigma$ term estimated from the nucleon mass through
(4) in full QCD with the Kogut-Susskind quark action is similarly
small $\sigma\approx 20$MeV.
For this action the quark mass obtained for
quenched and full QCD is consistent, being in the range
$\hat m\approx 2-3$MeV for both cases.

Origin of the small value of quark mass and that of the $\sigma$ term
in full QCD is not clear at present.
The discrepancy of quark mass between quenched and full QCD with
the Wilson quark action is also not understood.

\subsubsection{strange quark content within nucleon}

\begin{table}[t]
\setlength{\tabcolsep}{0.8pc}
\caption{Results for nucleon scalar matrix elements}
\label{tab:strange}
\begin{tabular}{lll}
\hline
  & Ref.~\cite{ktsigma}& Ref.~\cite{liulat}\\
\hline
\multicolumn{3}{l}{(a) $\langle N\vert {\bar u}u+
{\bar d}d\vert N\rangle$}\\
connected&2.52(6)&2.85(6)\\
disconnected&6.2(1.9)&3.79(13)\\
total&8.9(1.9)&6.6(2)\\
\hline
\multicolumn{3}{l}{(b) strangeness content}\\
$\langle N\vert {\bar s} s \vert N \rangle$ & 2.89(61)&1.98(9)\\
$m_s \langle N\vert {\bar s} s \vert N \rangle/m_N$ & 0.40(8)&0.27(1)\\
$y={2\langle N\vert {\bar s} s \vert N \rangle \over
\langle N\vert {\bar u} u + {\bar d} d \vert N
\rangle}$& 0.65(20)&0.60(4)\\
\hline
\end{tabular}
\vspace*{-6mm}
\end{table}

The nucleon matrix element of the strange
quark density $\langle N\vert {\bar s} s \vert N \rangle$
receives contribution only from the disconnected amplitude.
In Table~2 (b) we summarize recent results\cite{ktsigma,liulat}.  These
values are obtained by extrapolating
the valence quark mass to the chiral limit while keeping the strange quark
mass fixed at the physical value.
Although the strange quark density within nucleon is not directly
measurable by experiments, both the ratio
$m_s \langle N\vert {\bar s} s \vert N \rangle/m_N$, which naively
measures the strange quark contribution to the nucleon mass, and
the $y$ parameter appear quite large.

\begin{figure}[t]
\vspace*{-0.8cm}
\begin{center}
\leavevmode\psfig{file=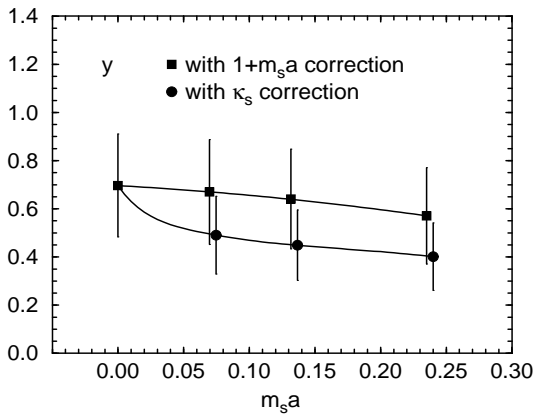,height=5.5cm,width=7.0cm,angle=-90}
\end{center}
\vskip -1.2cm
\caption{$y$ parameter as functions of $m_s a$ with $1+m_s a$ or
$\kappa_S(m_s a)$ factor for strange quark density.}
\label{fig:y}
\vspace{-6mm}
\end{figure}

Concerning this point Laga${\rm \ddot e}$ and Liu suggested\cite{lagae}
that the quark mass correction factor $1+m_qa$\cite{lepage91},
derived for valence quarks nearly on the mass shell, may not adequately
account for $O(m_qa)$ effects for sea quarks in disconnected
diagrams,  which are mostly off the mass-shell states.
To examine this possibility they calculated the ratio $\kappa_S(m_qa)$ of
the triangle
diagram for the operator ${\bar q}q$ with two external gluons
at zero momenta on the lattice and in the continuum
in one-loop perturbation theory.

In Fig.~\ref{fig:y} we compare the $m_s a$ dependence of the $y$ parameter
for the data of the
Kyoto-Tsukuba group corrected by the $1+m_s a$ factor and by the
Laga${\rm \ddot e}$-Liu $\kappa_S$ factor.
With the $\kappa_S$ factor the $y$ parameter is reduced by about
30\% in the region of strange quark mass $m_s a \approx 0.1$,
and $y$ values of the Kyoto-Tsukuba
and Kentucky groups become $y=0.46(14)$ and $0.42(3)$.
This suggests a possible presence of large $O(m_q a)$ effects for the
nucleon matrix element of strange quark density.

\subsubsection{baryon mass splitting and reduced matrix elements
$F_S$ and $D_S$}

\begin{figure}[t]
\vspace*{-0.4cm}
\begin{center}
\leavevmode\psfig{file=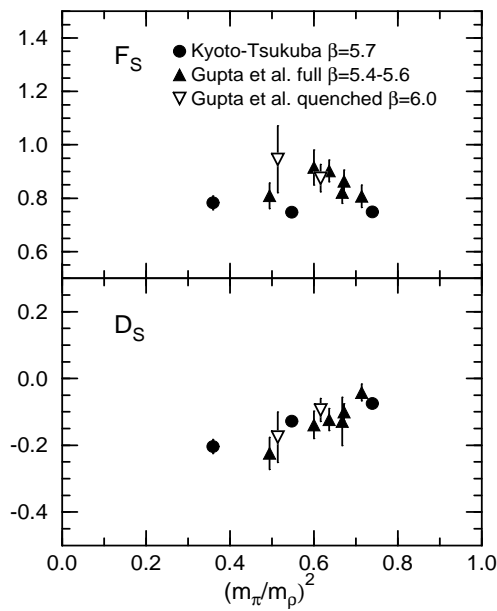,height=8.0cm,width=6.5cm}
\end{center}
\vskip -1.2cm
\caption{Reduced scalar matrix elements $F_S$ and $D_S$.
For reference see Table 1.}
\label{fig:reduce_scalar}
\vspace{-6mm}
\end{figure}

For quenched QCD lattice estimates for the $\sigma$ term
reasonably agree with the phenomenological value.
Within the first-order flavor $SU(3)$ breaking formula
(\ref{eq:sigma0}--\ref{eq:firstorder}), a large value of $y$ is not
consistent with this result unless the mass splitting of the baryon octet,
or equivalently the matrix element
$\langle N\vert\overline{u}u+\overline{d}d-2\overline{s}s\vert N\rangle$
$\approx$
$\langle N\vert\overline{u}u+\overline{d}d\vert N\rangle_{\rm conn.}$, is
small.

To examine this point we recall the definition of
reduced matrix elements $F_S$ and $D_S$ given by
\begin{eqnarray}
\langle B \vert {\bar q} \lambda_8 q \vert B \rangle
&=&\ F_S {\rm Tr}(B^\dagger[\lambda_8,B]) \nonumber \\
&+& D_S {\rm Tr}(B^\dagger\{\lambda_8,B\}).
\label{eq:FsDs}
\end{eqnarray}
To first order in flavor $SU(3)$ breaking, we have the relations
$F_S=(M_\Xi-M_N)/2(m_s-{\hat m})$ and
$D_S=(M_\Xi+M_N-2M_\Sigma)/2(m_s-{\hat m})$, from which we find
experimentally that
\begin{equation}
F_S^{\rm exp.}=1.52,\ \ \ \ \ \ D_S^{\rm exp.}=-0.518.
\label{eq:FsDs_exp}
\end{equation}

These values are to be compared with lattice estimates
obtained from the relations,
\begin{eqnarray}
F_S&=&{1\over 2}\langle N \vert {\bar u} u \vert N \rangle_{\rm conn.}
\nonumber \\
D_S&=&{1\over 2}\langle N \vert {\bar u}u -2 {\bar d} d
\vert N \rangle_{\rm conn.}
\label{eq:FsDs_lat}
\end{eqnarray}
where only connected contributions are kept since disconnected
contributions are to be ignored to leading order of flavor
symmetry breaking.

Lattice results from the studies in Table~\ref{table:tabone} are plotted
in Fig.~\ref{fig:reduce_scalar}.
Previous estimates\cite{romeold,wuppertalold}
are similar with results in this figure.
While there may be some trend of increase
toward the chiral limit, both matrix elements are
small compared to experiments.  It should be added that the ratio
$F_S/D_S$ is consistent, however.

One can estimate the baryon mass splitting from results for the reduced
matrix elements.  At $\beta=5.7$, for example, one finds
$M_\Xi-M_N=177(8)$MeV and $M_\Sigma-M_N=114(5)$MeV\cite{ktsigma}
as compared to the experimental values 379MeV and 254MeV.
On the other hand,
a recent spectrum calculation at $\beta=6.0$ on a $32^3\times 64$
lattice reported $M_\Xi-M_N=300(27)$MeV and $M_\Sigma-M_N
=185(17)$MeV\cite{guptaC}, which show a much better agreement with
experiments.

Systematic analyses calculating both the
scalar matrix elements and the baryon mass splitting within the
same simulation have not been carried out so far.  These are needed to
elucidate the origin of the problem in these quantities, especally to see
whether it can be ascribed to a large violation of scaling.

\begin{figure}[t]
\vspace*{-0.8cm}
\begin{center}
\leavevmode\psfig{file=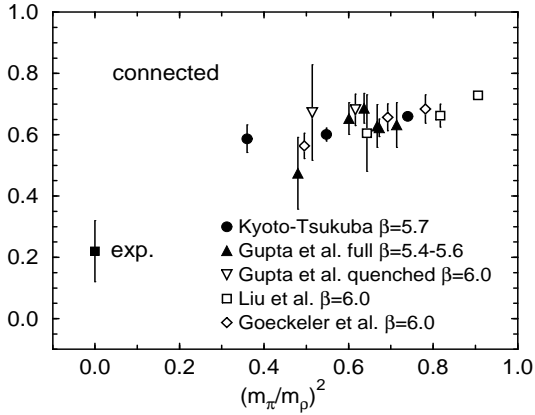,height=5.5cm,width=7.0cm,angle=-90}
\end{center}
\vskip -1.2cm
\caption{Connected contribution
$\Delta \Sigma_{\rm conn.} =(\Delta u + \Delta d)_{\rm conn.}$
of the axial vector matrix element
as a function of $(m_\pi/m_\rho)^2$.  For reference see Table 1.}
\label{fig:spin_conn}
\vspace{-6mm}
\end{figure}

\section{NUCLEON AXIAL VECTOR MATRIX ELEMENTS}

\subsection{Quark content of proton spin}

The forward matrix element of the axial vector current
${\bar q} \gamma_\mu \gamma_5 q$ measures the contribution
$\Delta q$ of a given quark flavor $q$ to the spin of the proton,
\begin{equation}
\langle P_s \vert {\bar q} \gamma_3 \gamma_5 q \vert P_s \rangle
= s\cdot \Delta q,
\label{eq:spin}
\end{equation}
where $\vert P_s \rangle$ is the proton state at rest with the spin
projection in the $z$ direction equal to $s/2$.
Recent experiments on polarized deep
inelastic lepton-nucleon scattering from
SMC\cite{smc} and E143\cite{e143} have confirmed the
earlier finding of EMC\cite{emc}
that the fraction of proton spin carried by quarks
has a small value $\Delta \Sigma = \Delta u +\Delta d +\Delta s \approx
0.2-0.3$
and that the strange quark contribution is non-vanishing and negative
$\Delta s \approx -0.1$.  These values are quite different from naive
expectations from quark models.   We tabulate the experimental
values and those from quark models in Table 3 (a) and (b).

Recent lattice results for the connected contribution
$\Delta \Sigma_{\rm conn.} =(\Delta u + \Delta d)_{\rm conn.}$
are compiled in Fig.~\ref{fig:spin_conn} where we used the
tadpole-improved renormalization factor given by\cite{tadpole}
\begin{equation}
Z_A\ =\ \left(1-{3K \over 4K_c}\right)
\biggl[1-0.31\alpha_{\overline {\rm MS}}({1 \over a})\biggr].
\label{eq:ZA}
\end{equation}
Lattice parameters for individual runs are summarized in Table 1.
For comparison we also plot the experimental value of SMC at
$(m_\pi/m_\rho)^2$ = 0.  The connected contribution alone is too large
to explain the small value of $\Delta\Sigma$.

\begin{figure}[t]
\vspace*{-0.8cm}
\begin{center}
\leavevmode\psfig{file=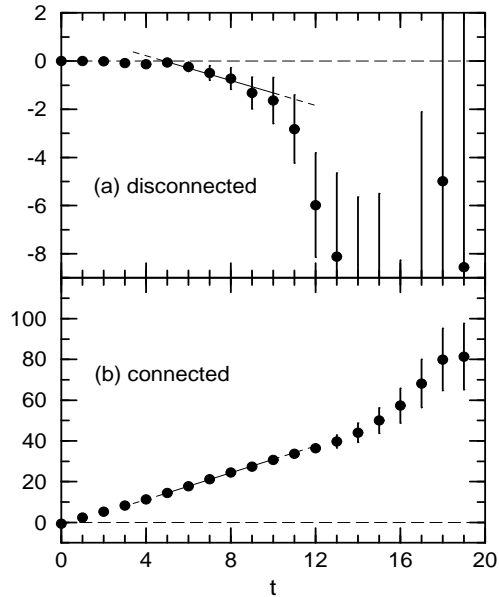,height=8.0cm,width=6.5cm}
\end{center}
\vskip -1.2cm
\caption{Ratio $R(t)$ for the (a) disconnected and (b) connected amplitudes
of the axial vector matrix element $\Delta u$
evaluated by the wall source method without gauge fixing\cite{lat94,ktspin}
for $K$=0.164 at $\beta$=5.7
on a $16^3\times 20$ lattice.  Solid lines are linear fits over
$5 \le t \le 10$.}
\label{fig:rakt}
\vspace{-6mm}
\end{figure}

\begin{figure}
\vspace*{-0.8cm}
\begin{center}
\leavevmode\psfig{file=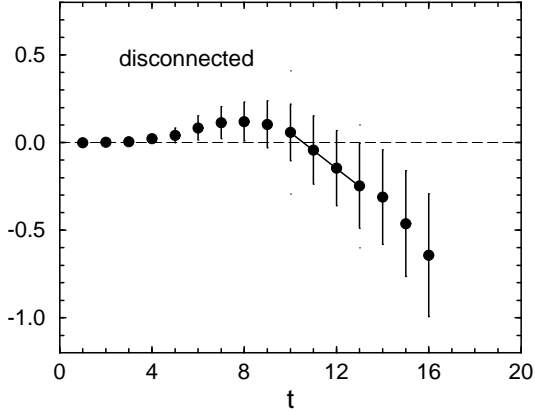,height=5.5cm,width=7.0cm,angle=-90}
\end{center}
\vskip -1.2cm
\caption{Ratio $R(t)$ for the disconnected amplitude
of the axial vector matrix element
$\Delta u$
evaluated by the $Z_2$ source method\cite{liulat,liuspin}
for $K$=0.154 at $\beta$=6.0
on a $16^3\times 24$ lattice.  Solid line is a linear fit over
$10 \le t \le 13$.}
\label{fig:raliu}
\vspace{-6mm}
\end{figure}

\begin{table}[t]
\setlength{\tabcolsep}{0.1pc}
\caption{Axial vector matrix elements from (a) experiments, (b)quark models
and (c) lattice calculations.}
\label{tab:efflue}
\begin{tabular}{lllll}
\hline
       & $\Delta u$ & $\Delta d$ & $\Delta s$ & $\Delta \Sigma$ \\
\hline
\multicolumn{5}{l}{(a) Experiment}\\
SMC\cite{smc} & 0.80(6) & -$0.46(6)$ & -$0.12(4)$ & 0.22(10) \\
E143\cite{e143} & 0.82(6) & -$0.44(6)$ & -$0.10(4)$ & 0.27(10) \\
\hline
\multicolumn{5}{l}{(b) Quark model} \\
non-rel. & 4/3 & -$1/3$ & 0 & 1 \\
rel. & 1.01 & -$0.25$ & 0 & 0.76 \\
\hline
\multicolumn{5}{l}{(c)Lattice results at $m_q=0$} \\
Ref.~\cite{ktspin}
& 0.638(54) & -$0.347(46)$ & -$0.109(30)$ & 0.18(10) \\
Ref.~\cite{liuspin}
& 0.79(11) & -$0.42(11)$ & -$0.12(1)$ & 0.25(12) \\
\hline
\end{tabular}
\vspace*{-0.5cm}
\end{table}

In Fig.~\ref{fig:rakt} we show results for the $R$ ratio (\ref{eq:ratio})
for the up quark contribution from Ref.~\cite{lat94,ktspin} where the
disconnected contribution (Fig.~\ref{fig:rakt}(a)) is obtained with the
method of wall source without gauge fixing.
Compared to the case of the scalar matrix element,
the data for the disconnected contribution are much noisier.
The negative value of the matrix element, not predictable
in quark models, is clearly observed, however.  A similar trend is also
seen in the result obtained with the $Z_2$ source shown in
Fig.~\ref{fig:raliu}\cite{liulat,liuspin}.

\begin{figure}[t]
\vspace*{-0.95cm}
\begin{center}
\leavevmode\psfig{file=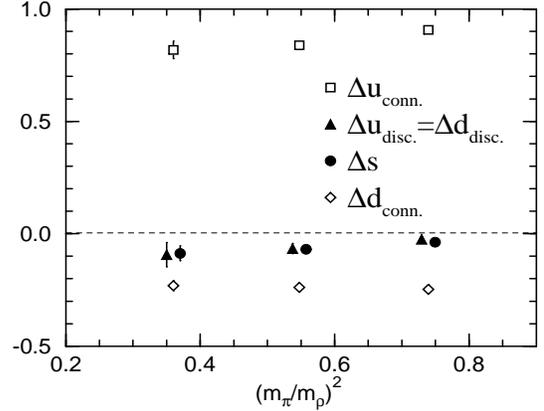,height=5.5cm,width=7.0cm,angle=-90}
\end{center}
\vskip -1.2cm
\caption{Axial-vector matrix elements for $u$, $d$ and $s$ quarks as
functions of $(m_\pi/m_\rho)^2$\protect\cite{ktspin}.
Strange quark mass is fixed to the
physical value.
}
\label{fig:deluds}
\vspace{-6mm}
\end{figure}

\begin{figure}[t]
\vspace*{-0cm}
\begin{center}
\leavevmode\psfig{file=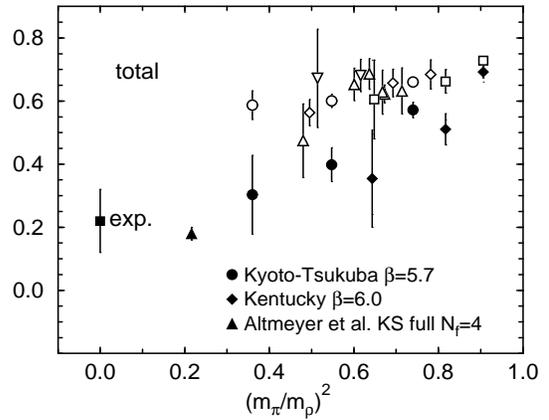,height=5.5cm,width=7.0cm,angle=-90}
\end{center}
\vskip -1.2cm
\caption{Total value of $\Delta \Sigma$
as a function of $(m_\pi/m_\rho)^2$.
For reference see Table 1.
Triangle is a result using the anomaly equation (5).
}
\label{fig:spin_total}
\vspace{-6mm}
\end{figure}

Values of the connected and disconnected contributions
obtained in ref. \cite{ktspin}
are plotted in Fig.~\ref{fig:deluds} for $u$, $d$ and $s$ quarks
separately as functions of $(m_\pi/m_\rho)^2$.
The magnitude of disconnected contributions is small for each quark.
However, the total amplitude
$\Delta\Sigma$ receives a negative disconnected contribution from
$u$, $d$ and $s$ quarks.  As a result the value of
$\Delta\Sigma$ is significantly
reduced from that obtained from connected contributions alone.
This is shown in Fig.~\ref{fig:spin_total} where filled symbols
are the total amplitude and open ones connected contributions
taken from Fig.~\ref{fig:spin_conn}. We observe that the total amplitude
points toward the experimental value for light quarks.

A more detailed
comparison is made in Table 3 where the total contribution
extrapolated to $m_q=0$ is listed
for each quark.  Lattice results are in reasonable agreement with
those from experiments.

\subsection{Isovector axial coupling $g_A$}

The $SU(3)$ reduced matrix elements for the axial vector current are
given by
\begin{eqnarray}
F_A&=&\frac{1}{2}\left(\Delta u\right)_{\rm conn.}\\
D_A&=&\frac{1}{2}\left(\Delta u - 2\Delta d\right)_{\rm conn.}.
\end{eqnarray}
Experimental values of these matrix elements are deduced using neutron
$\beta$ decay, which gives a very precise value of the isovector axial
coupling $g_A$,
\begin{equation}
g_A^{\rm exp.} = F_A+D_A=1.2573(28)
\label{eq:G_A_exp}
\end{equation}
and hyperon $\beta$ decay, which yields
\begin{equation}
F_A-D_A=-0.327(20).
\end{equation}
{}From the two estimates one finds for the ratio,
\begin{equation}
F_A/D_A=0.58(5)
\end{equation}

\begin{figure}[t]
\vspace*{-0.5cm}
\begin{center}
\leavevmode\psfig{file=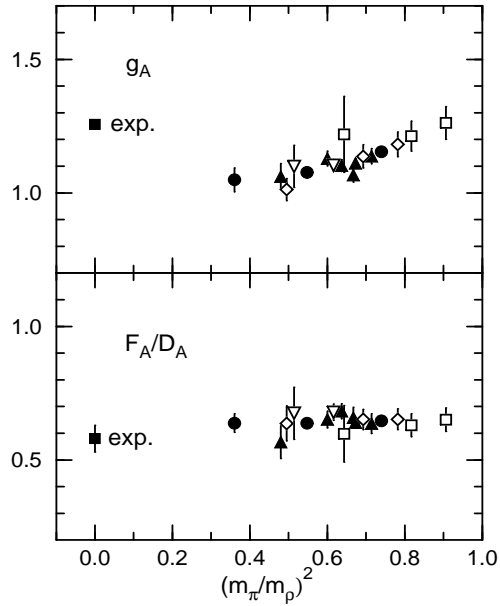,height=8.0cm,width=6.5cm}
\end{center}
\vskip -1.2cm
\caption{Isovector axial coupling $g_A$ and $F_A/D_A$
as functions of $(m_\pi/m_\rho)^2$.  For the meaning of symbols
see Fig. 8.}
\label{fig:G_A}
\vspace{-6mm}
\end{figure}

In Fig.~\ref{fig:G_A} we plot recent lattice results for $g_A$ and the
ratio $F_A/D_A$.  We see that lattice data for $g_A$ are consistently
smaller than the experimental value by about 20-30 $\%$ toward light
quarks, while the ratio $F_A/D_A$ is consistent.

We remark that a previous result\cite{wuppertalold} for a
$\sqrt{3}$-blocked quark action shows a simiar trend if a
tadpole-improved renormalization
factor appropriate for the action is employed.

The disagreement of $g_A$ represents another problem with flavor
non-singlet nucleon matrix elements.
It does not seem to arise from the effect of
quenching since full QCD estimates plotted by upward
triangles\cite{guptaB} do not deviate from quenched results.
Effects of scaling violations also seem small since
results from two values of $\beta$ at 5.7\cite{ktspin} and at
6.0\cite{guptaB,liuaxial,desy} are in agreement.
One may suspect a possible ambiguity in the choice of the renormalization
constant $Z_A$, especially because the ratio $F_A/D_A$ agrees
with experiments.
However, the same renormalization factor enters in a calculation of the
pion decay constant $f_\pi$, for which quenched lattice results at
$\beta=5.7-6.0$ agree with the experimental value within
about $10$\%.
Furthermore a non-perturbatively  determined value of $Z_A$
obtained at $\beta$ = 6.0 and $K$= 0.1515 and 0.153 is quite consistent
with the perturbative result (\ref{eq:ZA})\cite{zz}.

\section{HADRON SCATTERING LENGTH}

The theoretical basis for a lattice calculation of hadron scattering
lengths is the formula proven by
L${\rm \ddot u}$scher\cite{refluscher}; it
relates the $s$-wave scattering length $a_0$
to the energy level $E$ of the lightest two-hadron state with vanishing
spatial momentum in a cubic box of length $L$ according to
\begin{equation}
E-2m =\ -{4\pi a_0 \over m L^3}
(1+c_1{a_0 \over L}+c_2{a_0^2 \over L^2})\ +\ O({1 \over L^6})
\label{Luscher}
\end{equation}
with $c_1=-2.837297$ and $c_2=6.375183$.
The technical difficulty of calculating general four-point functions
(including disconnected ones) needed for an extraction of the energy level
$E$ was overcome by the wall source method without gauge fixing, and
$\pi$-$\pi$, $\pi$-$N$, $K$-$N$ and $N$-$N$ scattering lengths have been
calculated in quenched QCD\cite{pipisct,NNsct,hadsct}.
While no new simulations have been carried out since
{\it LATTICE 93},
two theoretical studies aiming at a deeper understanding of the
lattice calculation of hadron scattering lengths have been made
(see also Refs.~\cite{fiebig,gottlieb} for related investigations).

\subsection{$\pi$-$\pi$ scattering length}

Chiral perturbation theory suggests that quantities calculated in quenched
QCD often develop spurious infrared divergences in the chiral limit,
which are not present in full QCD\cite{qchiral}.
Bernard and Golterman\cite{BG}
investigated the effect of quenching in the
energy level $E$ of the two-pion state using quenched chiral
perturbation theory technique, extending a previous work on the
subject\cite{BGLSU}.  They found that enhanced finite-volume corrections
 of order $L^0=1$ and $L^{-1}$ occur
in $E$ at one loop due to the double pole in the $\eta'$ propagator.
This implies that it is not consistent to use L${\rm \ddot u}$scher's
formula in quenched QCD.

On the other hand, a quenched lattice calculation of $\pi$-$\pi$
scattering lengths
with the Kogut-Susskind quark action\cite{pipisct} yielded results
consistent with the prediction of current algebra and PCAC, which is
known to hold for the Kogut-Susskind case\cite{lanlpipi}.
They suggest that this agreement is due to an accidental choice of
lattice parameters such that the enhanced
finite-volume corrections are not large for the values of $m_\pi L$ of the
simulation.

\begin{figure}[t]
\vspace*{-0.8cm}
\begin{center}
\leavevmode\psfig{file=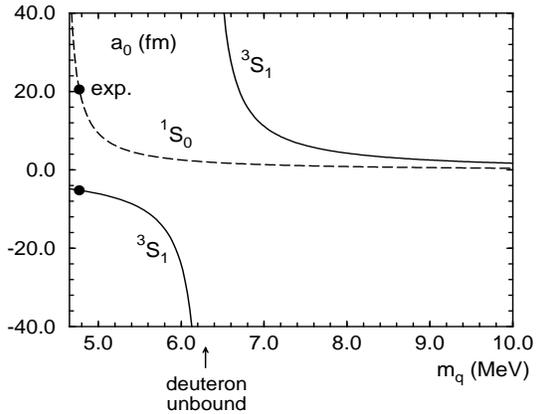,height=5.5cm,width=7.0cm,angle=-90}
\end{center}
\vskip -1.2cm
\caption{$N$-$N$ scattering length $a_0$ calculated by
a one-boson exchange model for ${}^1 S_0$ and ${}^3 S_1$
channels as functions of quark mass $m_q$.
The physical point corresponds to $m_q$ = 4.8 MeV.}
\label{fig:scatter}
\vspace{-6mm}
\end{figure}

\subsection{$N$-$N$ scattering length}

The nucleon-nucleon scattering length is a fundamental parameter
characterizing the low-energy properties of the nuclear force.
Experimentally they
are well determined and are known to have large values,
\begin{eqnarray}
a_0({}^1 S_0)\ &=& +20.1(4)\ {\rm fm},\nonumber \\
a_0({}^3 S_1)\ &=& -5.432(5)\ {\rm fm}.
\label{nnsl}
\end{eqnarray}
Since there are no constraints from chiral symmetry for $N$-$N$
scattering, explaining these values from first principles of lattice
QCD is an
interesting dynamical problem.  However, a number of obstacles exist
toward realistic calculations.  A very large lattice is needed to
apply L\"uscher's formula in order to suppress $O(L^{-6})$ corrections
neglected in (\ref{Luscher}).
Extraction of the scattering length in the ${}^3 S_1$ channel
requires a calculation of the lowest scattering state orthogonal to the
ground state, which is the deuteron bound state.

A possible strategy in this situation would be
to examine the quark mass dependence of the scattering lengths starting
from a heavy quark where simulations are much easier. For this approach
we need to understand the behavior of scattering lengths for
heavy quarks.

For heavy quarks the $N$-$N$ interaction becomes shorter ranged since
pions exchanged between nucleons are heavier.  This will force the
nucleons out of the attractive well of the potential, which suggests
that the deuteron becomes unbound as the quark mass increases.

A closer examination of this possibility was
made in Ref.~\cite{hadsct}, taking a phenomenological model of one-boson
exchange potentials\cite{holinde} and
varying hadron masses in the model as a function of quark mass according
to lattice data.
The results for the $N$-$N$ scattering lengths are plotted in
Fig.~\ref{fig:scatter}.
The divergence of the scattering
length for the ${}^3 S_1$ channel taking place at $m_q$ = 6.3 MeV signals
unbinding of the deuteron.  For $m_q > $ 6.3 MeV, both $a_0({}^1 S_0)$
and $a_0({}^3 S_1)$ are positive and rapidly decrease to have values of
order 1 fm.  Also the value of $a_0({}^3 S_1)$ is larger than that for
$a_0({}^1 S_0)$ for heavy quark indicating a stronger attraction in
the ${}^3 S_1$ channel.

\begin{figure}[t]
\vspace*{-0.8cm}
\begin{center}
\leavevmode\psfig{file=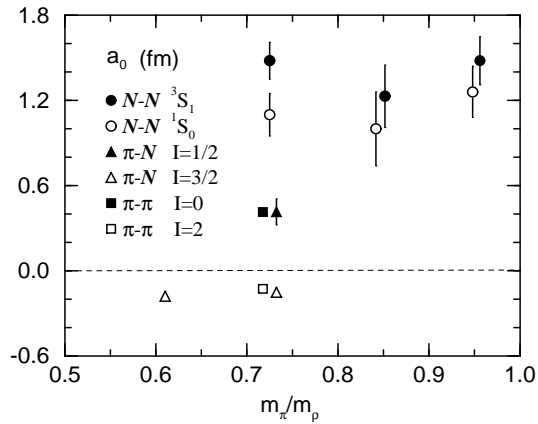,height=5.6cm,width=7.0cm,angle=-90}
\end{center}
\vskip -1.2cm
\caption{$N$-$N$ scattering length in physical units at $\beta$=5.7 on a
$20^4$
lattice.  Also shown are the $\pi$-$\pi$ and $\pi$-$N$ scattering lengths
at the same $\beta$ on an $12^3\times 20$ lattice.}
\label{fig:pp}
\vspace{-8mm}
\end{figure}

These are precisely the features found in a lattice calculation
of $N$-$N$ scattering lengths\cite{NNsct,hadsct}.  The results are shown
in Fig.~\ref{fig:pp} together with those for $\pi$-$\pi$
and $\pi$-$N$ cases for comparison.

\section{CONCLUSION}

In the last two years, a substantial progress
has been made in the lattice QCD study
of the flavor singlet nucleon matrix elements.  Applications of efficient
calculational techniques for handling disconnected quark loop amplitudes
revealed significant sea quark contributions in the flavor singlet
scalar and axial vector matrix elements.  Adding sea and valence
quark contributions the
total magnitude of these matrix elements are comparable to experimental
estimates.
Although the nature of systematic errors such as
scaling violations and quenching effects need to be understood more
precisely, the encouraging results found so far point
toward an eventual resolution of the problems related to the flavor
singlet matrix elements within lattice QCD.

In the course of the studies it has become apparent that current lattice
results for flavor non-singlet scalar and axial vector matrix elements
differ from experiments beyond statistical errors. A systematic scaling
study appears needed to pinpoint the origin of the problem.

We have also discussed some recent work in the study of hadron scattering
lengths.  The model result that the deuteron becomes unbound for a quark
mass only slightly larger than the physical value suggests that much
interesting physics can be done on the subject of $N$-$N$ scattering.
At the same time the
possibility of a severe quenching error pointed out for
the $\pi$-$\pi$ case signifies the need of caution in the use of quenched
QCD.

Study of flavor singlet physics is a young subject in lattice QCD.
Results obtained so far should serve as a stepping stone for further
development in this area in the years to come.

\vspace{0.3cm}
\noindent
${\rm \bf Acknowledgements \hfill}$

I would like to thank C. Bernard, S. Dong, M. Fukugita, Y. Kuramashi,
K. Liu, G. Schierholz and A. Ukawa for correspondence and discussions.
Valuable suggestions of A. Ukawa on the manuscript are gratefully
acknowledged.  This work is supported in part by the Grant-in-Aid of the
Ministry of Education (No. 05640363).

\end{document}